\documentclass[journal=jpcafh,manuscript=article]{achemso}
\setkeys{acs}{usetitle = true}

\usepackage[version=3]{mhchem} 
\usepackage[T1]{fontenc}       
\usepackage{color}
\usepackage[english]{babel}
\usepackage{rotating}

\author{Thanja Lamberts}
\email{lamberts@theochem.uni-stuttgart.de}
\phone{ +49 (0)711 685 64833}
\affiliation[University of Stuttgart]
{Institute for Theoretical Chemistry, University of Stuttgart, Pfaffenwaldring 55, 70569 Stuttgart, Germany}
\author{Johannes K\"astner}
\affiliation[University of Stuttgart]
{Institute for Theoretical Chemistry, University of Stuttgart, Pfaffenwaldring 55, 70569 Stuttgart, Germany}

\title[\ce{H + H2S -> H2 + HS}]
  {Tunneling Reaction Kinetics for the Hydrogen Abstraction Reaction \ce{H + H2S -> H2 + HS} in the Interstellar Medium}

\begin{document}

\begin{tocentry}
\includegraphics[width=8.5cm]{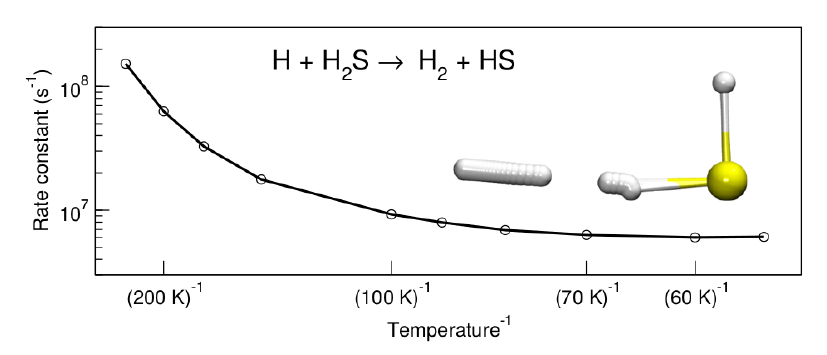}

\end{tocentry}

\begin{abstract}
\noindent The hydrogen abstraction reaction between H and \ce{H2S}, yielding HS and \ce{H2} as products, has been studied within the framework of interstellar surface chemistry. High-temperature rate constants up to 2000~K are calculated in the gas phase and are in agreement with previously reported values. Subsequently low-temperature rate constants down to 55~K are presented for the first time that are of interest to astrochemistry, i.e., covering both bimolecular and unimolecular reaction mechanisms. For this, a so-called implicit surface model is used.  Strictly speaking, this is a structural gas-phase model in which the restriction of the rotation in the solid state is taken into account. The calculated kinetic isotope effects are explained in terms of difference in activation and delocalization. All rate constants are calculated at the  UCCSD(T)-F12/cc-VTZ-F12 level of theory. Finally, we show that the energetics of the reaction is only affected to a small extent by the presence of \ce{H2O} or \ce{H2S} molecular clusters that simulate an ice surface, calculated at the MPWB1K/def2-TZVP level of theory.
\end{abstract}

\section{Introduction}

Many interstellar sulfourous compounds have been detected in the gas phase in the past decades\citep{uni-koeln}. In the ices present in the dark regions of the interstellar medium (ISM), on the other hand, only OCS and tentatively \ce{SO2} are identified \citep{Palumbo:1997, Boogert:1997}. Even so, \ce{H2S} has been put forward as a key species in interstellar ice chemistry, and has even been suggested to be responsible for the sulphur depletion in the gas phase in the dense (dark) regions \citep{Esplugues:2014,Holdship:2016,Danilovich:2017}. Experimentally, it has been shown that a plethora of surface reactions with \ce{H2S} as a reactant can take place \citep{Garozzo:2010,Jimenez:2011,Chen:2015,Kanuchova:2017}. Most of these reactions, however, are induced by energetic processing through which, amongst others, the HS radical is formed. Here, we explore the non-energetic reaction between the hydrogen atom and the hydrogen sulfide molecule:
\begin{equation}
\ce{H + H2S} \longrightarrow \ce{H2 + HS}\;. \tag{R1}\label{R1}
\end{equation}
It is exothermic by about {58} kJ/mol, contrary to the analogous reaction between H and \ce{H2O} which is endothermic by about {62} kJ/mol. Reaction \ref{R1} has been proposed to take place as a surface reaction on interstellar ice-covered grains. Indeed it is incorporated specifically in several recent astrochemical surface models that also include the low-temperature regime (10--20~K) in dense molecular clouds \citep{Woods:2015,Holdship:2016,Vidal:2017}. The two main surface reaction mechanisms are the Langmuir--Hinshelwood (LH) and Eley--Rideal (ER) processes. The LH mechanism assumes two physisorbed reactants that diffuse on the surface, subsequently meet each other, and then form an encounter complex. This encounter complex may decay and lead to a reaction, the rate constant of which is unimolecular. In the ER process on the other hand, a physisorbed species reacts directly with a reactant that impinges from the gas phase. In this case, bimolecular rate constants are required. For a complete understanding of surface astrochemistry, both rate constants are needed.

Furthermore, the title reaction can be of importance in determining the final deuteration fraction of hydrogen sulfide in the ISM. The \ce{HDS}/\ce{H2S} and \ce{D2S}/\ce{H2S} ratios are brought about via a series of abstraction and addition reactions, listed below, with X = H or D:
\begin{align}
 \ce{X + H2S} &\longrightarrow \ce{HX + HS} \tag{R2a}\label{R2a} \\
 \ce{X + HS} &\longrightarrow \ce{HXS} \tag{R2b}\label{R2b} \\
 \ce{X + HDS} &\longrightarrow \ce{XH + DS} \tag{R2c}\label{R2c} \\
 \ce{X + HDS} &\longrightarrow \ce{XD + HS} \tag{R2d}\label{R2d} \\
 \ce{X + DS} &\longrightarrow \ce{DXS} \tag{R2e}\label{R2e} \\
 \ce{X + D2S} &\longrightarrow \ce{DX + DS} \;.\tag{R2f}\label{R2f}  
\end{align}
In the ISM the \ce{HDS}/\ce{H2S} ratio has been determined to be 0.1 in the protostar IRAS 16293 \citep{Dishoeck:1995}, and upper limits of $\sim$10$^{-3}$ were found in several hot cores\citep{Hatchell:1999}. These, as well as the recently reported cometary value of $1.2\times10^{-3}$ in 67P/Churyumov-Gerasimenko\citep{Altwegg:2017} are significantly higher than the cosmic D/H ratio of $10^{-5}$ \citep{Piskunov:1997,Oliveira:2003}. This can be seen as a sign that indeed surface chemistry plays an important role, as outlined by \citet{Caselli:2012}.

Recently, a full-dimensional potential energy surface for the gas-phase reaction has become available with the use of neural networks \citep{Lu:2016} with which quantum and quasi-classical dynamics studies have been performed subsequently \citep{Qi:2017}. Although an exchange reaction is in principle also possible, it was found that the abstraction channel is favoured. Within the framework of combustion chemistry, the importance and non-Arrhenius behavior of reaction \ref{R1} has been realized \citep{HynesWine:2000} and gas-phase rate constants have been both measured and calculated in a temperature range of 200--2560~K \citep{Cupitt:1970,Kurylo:1971,Rommel:1972,Bradley:1973,Pratt:1977,Wojciechowski:1979,Nicholas:1979,Husain:1980,Roth:1982,Clyne:1983,Yoshimura:1992,Peng:1999,Kurosaki:1999,Huang:2000,Vandeputte:2009,Lu:2016}. 
Non-Arrhenius behavior is a sign of the importance of tunneling for the reaction. The barrier ($\sim$12.7 kJ/mol) is indeed quite high to be overcome thermally as the temperature decreases. Since tunneling becomes relatively more important at the low-temperature regime in which we are interested, it has to be taken into account if reaction rate constants are calculated. Note that this also has an effect on the relative rate constants depending if X = H or D for the reaction sequence \ref{R2a}--\ref{R2f} because of the mass-dependency of tunneling.

Here, we are interested primarily in surface reaction rate constants. Therefore we will use the available literature on high-temperature gas-phase rate constants as a reference and subsequently expand the results to low-temperatures in an implicit surface model, elaborate on the isotope effects, and finally show the influence of \ce{H2S} and \ce{H2O} molecules in the direct vicinity of the reactants. Reaction rate constants are calculated at temperatures down to {55~K} with the use of instanton theory that indeed accounts for tunneling inherently. Finally, we will comment on the astrochemial implications and summarize the main findings of this study.

\section{Computational details}

\subsection{Electronic structure}

The core system is small, 4 atoms and 19 electrons, and can thus be treated well with highly accurate methods. Here, in particular, we made use of (unrestricted) coupled-cluster theory with singles, doubles, and perturbative triple excitations (UCCSD(T)) \citep{Knowles:1993, Knowles:err, Deegan:1994} and explicitly-correlated geminal functions (UCCSD(T)-F12)\citep{Adler:2007, Knizia:2009} employing a restricted Hartree--Fock (RHF) reference function and the cc-pVTZ-F12\citep{Peterson:2008} basis set as has been done in the work of \citet{Lu:2016}. The geometry optimizations are performed using DL-find\citep{Kaestner:2009} and Molpro 2012\citep{MOLPRO} within the Chemshell\citep{Sherwood:2003, Metz:2014} framework. 

In order to test the influence of nearby molecules and clusters on the reactivity, we made use of density functional theory (DFT) with Gaussian 09 \citep{g09}. {The functional MPWB1K\citep{Zhao:2004} and basis set def2-TZVP \citep{Weigend:1998} have been employed after comparing the performance to that of single-point energies at the UCCSD(T)-F12/cc-VTZ-F12 level.} The purpose of these clusters is to determine the range of activation energies brought about by neighboring molecules, \emph{i.e.}, mimicking a surface. The barrier height, the reaction energy, and four dimer interaction energies have been benchmarked and are listed in Table~\ref{benchmark} both excluding zero-point energy (ZPE), $V$, and including ZPE, $E$. The small energy differences for the activation and interaction energies show that the functional is capable of describing both the cluster structure and the barrier region, both of which are crucial here. 

{Visual Molecular Dynamics version 1.9.2\citep{VMD} and Grace version 5.1.23 \citep{Grace} were used for visualization.}

\begin{table}[h]
  \caption{Comparison of the activation, reaction, and interaction energies in kJ/mol between DFT (MPWB1K/def2-TZVP) and UCCSD(T)-F12/cc-VTZ-F12}
  \label{benchmark}
  \begin{tabular}{lrrrrrr}
    \hline
						& \multicolumn{3}{c}{MPWB1K}	& \multicolumn{3}{c}{UCCSD(T)-F12} \\
						& $V$ & ZPE & $E$ & $V$ & ZPE & $E$ \\
    \hline
    Activation energy\textsuperscript{\emph{a}}	& 15.4 & -2.4 & 13.0 & 15.8\textsuperscript{\;\;} & -3.1 & 12.7 \\
    Reaction energy\textsuperscript{\emph{a}} 	& -49.9 & 2.6 & -47.3 & -57.3\textsuperscript{\;\;} & 2.6 & -54.7 \\
    Interaction energy \ce{H2O}-\ce{H2O} 	& -22.9	& 8.9	& -14.0	& -21.0\textsuperscript{\emph{b}}	& 	& 	\\
    Interaction energy \ce{H2S}-\ce{H2S} 	& -5.8	& 4.3	& -1.5	& -6.9\textsuperscript{\emph{b}}	& 	& 	\\
    Interaction energy \ce{H2S}-\ce{H2O} 	& -11.8	& 5.9	& -5.9	& -11.3\textsuperscript{\emph{b}}	& 	& 	\\
    Interaction energy \ce{H2S}-\ce{H2O} 	& -11.9	& 5.7	& -6.2	& -12.3\textsuperscript{\emph{b}}	& 	& 	\\
    \hline
  \end{tabular}\\
  \textsuperscript{\emph{a}} with respect to separated reactants\\
  \textsuperscript{\emph{b}} UCCSD(T)-F12 single point energies of geometries calculated at the MPWB1K/def2-TZVP level
  \end{table}

\subsection{Rate constants}
Rate constants have only been calculated for the core system on UCCSD(T)-F12 level with instanton theory \citep{Langer:1967,Miller:1975,Callan:1977,Coleman:1977}, \emph{i.e.}, for the four-atomic system. 
{Instanton theory, also known as imaginary-$F$ method or harmonic quantum transition state theory, takes quantum effects of atomic movements into account by statistical Feynman path integral theory. Partition functions of both the reactant state and the transition-state-equivalent, the instanton path, are obtained by a steepest-descent approximation to the phase space integrals. The computationally required steps are twofold: (1) The discretized Feynman path has to be optimized to find the instanton, a first-order saddle point in the space of closed Feynman paths. This was performed with a modified Newton--Raphson method.\citep{Rommel:2011, Rommel:2011b} (2) The calculation of Hessians of the potential energy at all images of the Feynman path to evaluate the rate constant. For more details on instanton theory we refer to a recent review.\cite{Kaestner:2014} Instanton theory is generally considered to be more accurate than one-dimensional tunneling corrections, like Eckart or Bell corrections. Moreover, especially at low temperature it can be expected to be more accurate than small-curvature, large-curvature, or similar tunneling corrections as a result of the temperature-dependent optimization of the tunneling path.}
The paths were discretized to 60 images and have been calculated between 240 and 55~K. Such a path represents the instanton and is equivalent to the tunneling path with the highest statistical weight. An example is depicted in Fig.~\ref{instpath}. It spans the barrier region at temperatures close to the crossover temperature ($T_\text{c}$, see below) and extends towards the reactant region at low temperature. The crossover temperature is defined as 
\begin{equation} T_\text{c} = \frac{\hbar \omega_b}{2\pi k_\text{B}} \end{equation} 
where $\omega_b$ is the absolute value of the imaginary frequency at the transition state, $\hbar$ Planck's constant divided by $2\pi$ and $k_\text{B}$ Boltzmann's constant. $T_\text{c}$ indicates the temperature below which tunneling dominates the reaction. 

\begin{figure}[h]
  \includegraphics[width=8cm]{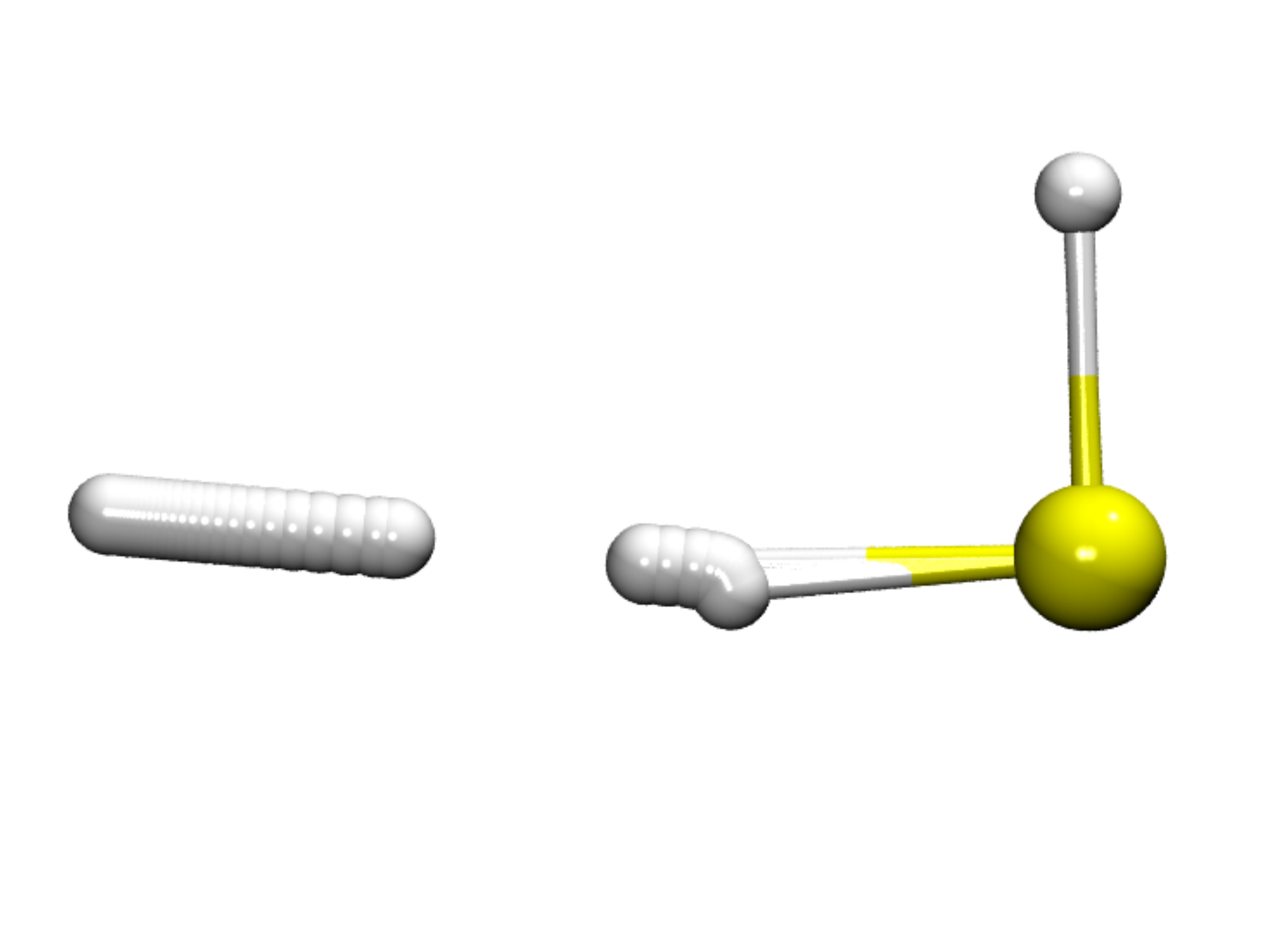}
  \caption{Instanton path calculated at 55~K.}
  \label{instpath}
\end{figure}

{Both bimolecular and unimolecular rate constants have been calculated. The former have been calculated with reduced instanton theory above $T_\text{c}$ \citep{Kryvohuz:2011, McConnel:2017b} and with canonical instanton theory below $T_\text{c}$, choosing parameters equivalently to previous work.\citep{Rommel:2011, Rommel:2011b}} Convergence is reached when all components of the nuclear gradient are smaller than $1\times10^{-8}$ atomic units. The gradient here is the derivative of the effective energy of the instanton with respect to the mass-weighted atomic coordinates. Rotational and translational partition functions were approximated by their classical analogues {(J-shifting approximation)}. For the reactant partition function in the bimolecular case, the product of the partition functions of the separated reactants was used. For unimolecular cases, the partition function of the encounter complex is employed. Rather than using the available PES \citep{Lu:2016} we make use of on-the-fly instanton calculations. This choice has been made since \citet{Lu:2016} state that the coordinates involving the attaching H atom were  sampled in relatively sparse grids. We are, however, also interested in unimolecular rate constants that represent the decay of the encounter complex H$\cdot\cdot\cdot$\ce{H2S} for which exactly this part of configuration space would be very important.

For rate constants relevant for surface reactions, several effects of the ice surface need to be taken into account. Thermal equilibrium is assumed at all stages of the calculation and thus excess heat is removed instantaneously. The structural effects are to a certain degree tested with the small cluster study. Finally, the restriction of the rotation of the reactants on a surface must be included as well. This can be mimicked in what is strictly speaking a gas-phase model, through modification of the translational and rotational partition functions. For the unimolecular case, the rotational partition function is assumed to be constant during the reaction, because rotational motions are suppressed in the reactant as well as in the transition state. For bimolecular rate constants, only translation of the H atom is considered, and both rotations and translations of the \ce{H2S} molecule are suppressed. This approach is referred to as ``implicit surface model''\citep{Meisner:2017}. Only in the gas phase, the full rotational partition function including its symmetry factor \citep{Fernandez:2007} is included.

\section{Results and discussion}

\subsection{Gas-phase reaction}

\begin{figure}[h]
  \includegraphics[width=9cm]{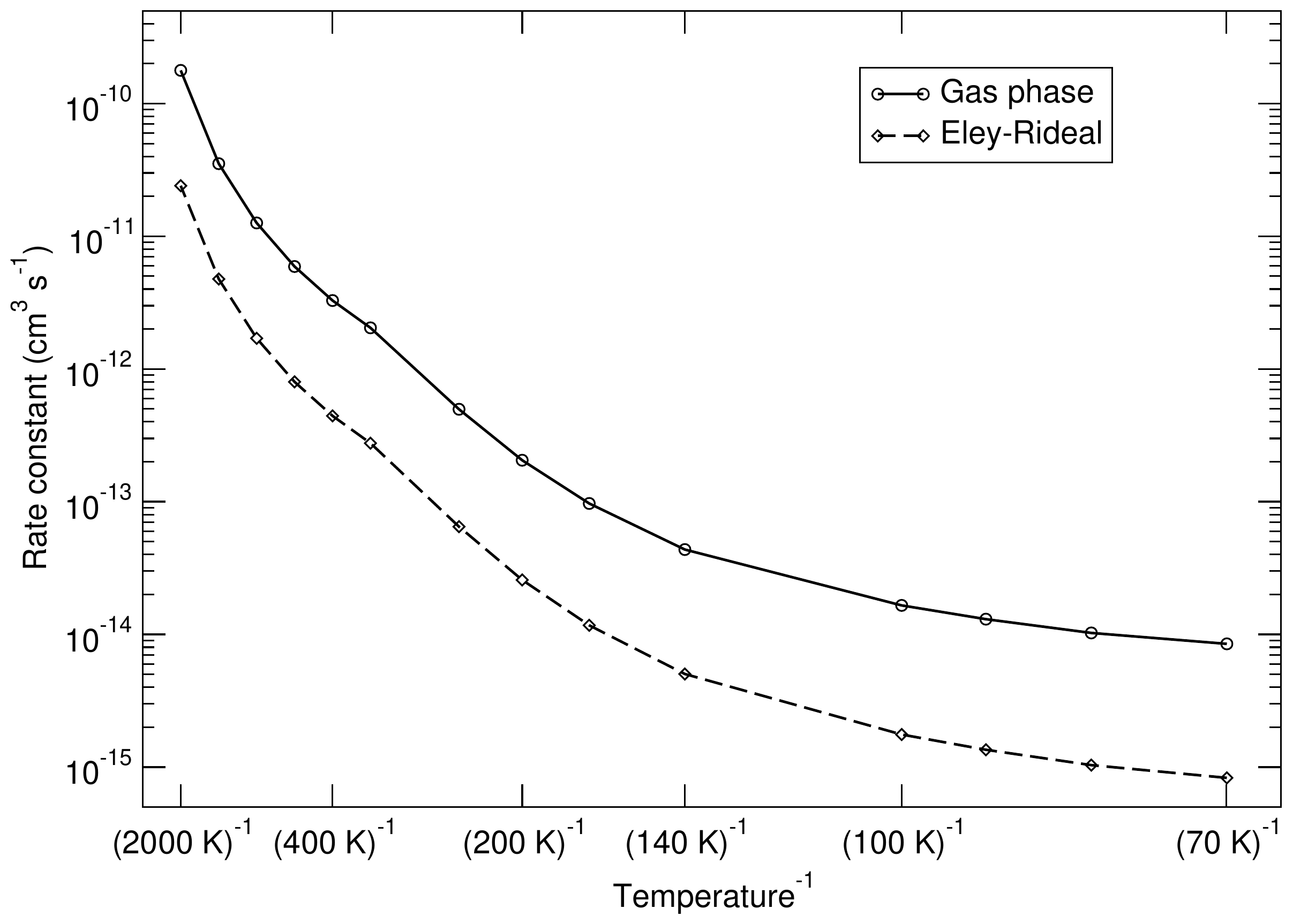}
  \caption{Bimolecular rate constants for the reaction \ce{H + H2S -> H2 + HS} in the temperature range of 2000 to 70~K.}
  \label{bimolgas}
\end{figure}

The calculated vibrationally adiabatic barrier of \ref{R1} given in Table~\ref{benchmark} amounts to a total of 12.7 kJ/mol with respect to the separated reactants. The crossover temperature is 285~K. Corresponding bimolecular rate constants are depicted in Fig.~\ref{bimolgas}. The high-temperature rate constants coincide with previous experimental and theoretical work \citep{Cupitt:1970,Kurylo:1971,Rommel:1972,Bradley:1973,Pratt:1977,Wojciechowski:1979,Nicholas:1979,Husain:1980,Roth:1982,Clyne:1983,Yoshimura:1992,Peng:1999,Kurosaki:1999,Huang:2000,Vandeputte:2009}. With a value of 1.8$\times 10^{-10}$ cm$^{-3}$ molecule$^{-1}$ s$^{-1}$ we are in close agreement with the experimental values of $\sim$1.5$\times 10^{-10}$ cm$^{-3}$ molecule$^{-1}$ s$^{-1}$ at 2000~K, and likewise our value of 2.3$\times 10^{-13}$ cm$^{-3}$ molecule$^{-1}$ s$^{-1}$ is close to the experimental value of $\sim$2$\times 10^{-13}$ cm$^{-3}$ molecule$^{-1}$ s$^{-1}$ at 200~K.
Note that these rate constants are calculated using $\sigma=2$. The precise values are listed in the Supporting Information. {The small bump in the curve at 333~K is due to the transition from reduced instanton theory to canonical instanton theory} Finally, the lower end of the temperature range is 70~K, because at even lower temperatures the tunneling energy drops below the total (separated) reactant energy and consequently the calculated rate constant is no longer reliable \citep{McConnel:2017}.

Fig.~\ref{bimolgas} also shows the influence of suppressing reactant rotation. At low temperatures this can lead up to an order of magnitude difference in the rate constant. The values for this bimolecular ER process are given in the Supporting Information. In the following Section the implicit surface model is used for all calculations.

\subsection{Isotope effects}

\begin{figure}[h]
  \includegraphics[width=9cm]{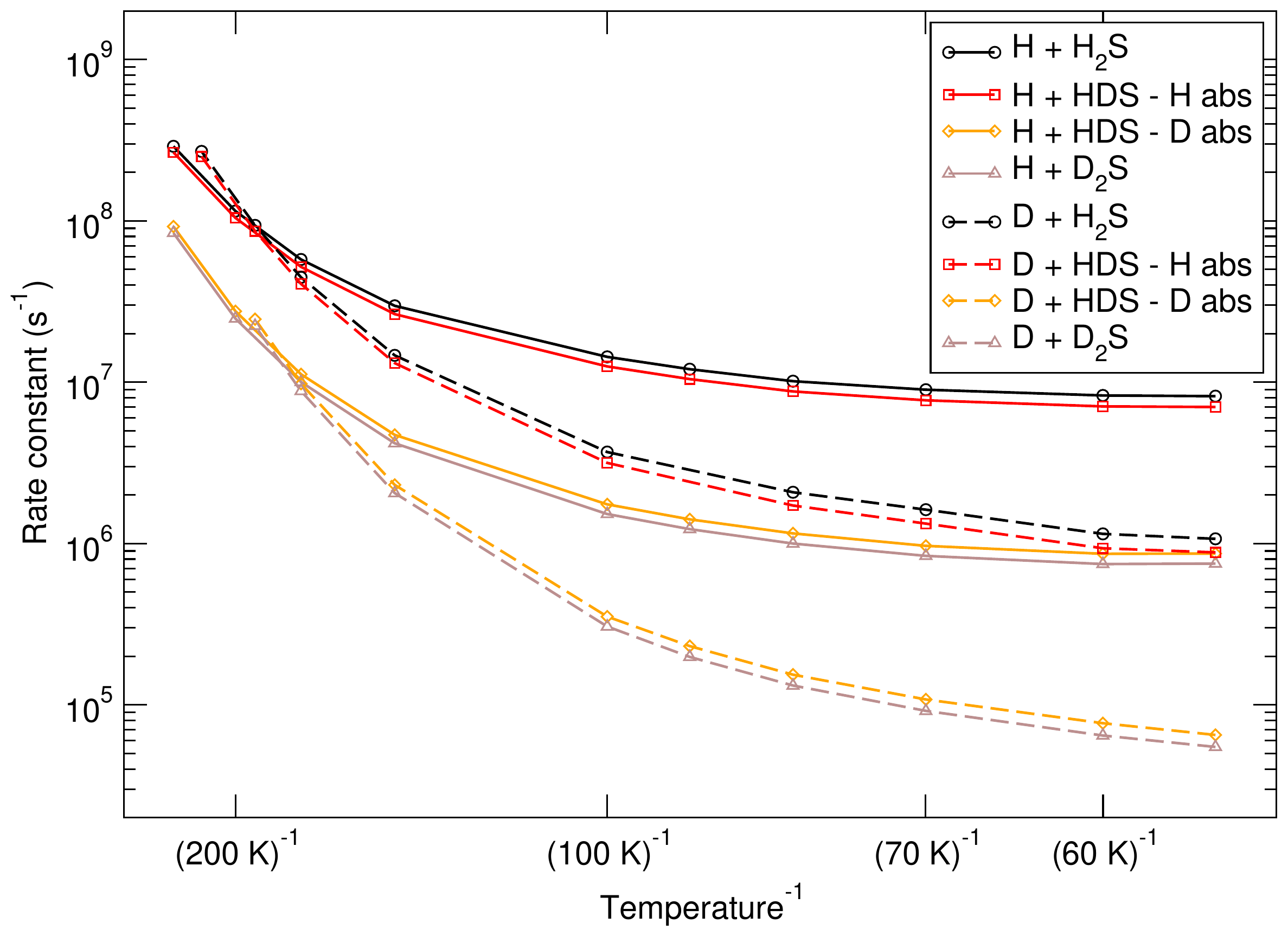}
  \caption{Unimolecular rate constants for the reaction \ce{H + H2S -> H2 + HS} and all isotope substitions in the temperature range of 240 to 55~K.}
  \label{unimol}
\end{figure}

The unimolecular rate constants of the decay of the pre-reactive complex H...\ce{H2S} for all eight isotope substitutions are depicted in Fig.~\ref{unimol}. The high-temperature behavior can be rationalized using the vibrationally adiabatic activation energies ($E_\text{uni, act.}$) summarized in Table~\ref{unimolbar}. The largest rate constants in the high-temperature regime are for those reactions with the lowest barriers. The influence of substituting a protium for a deuterium atom on low-temperature rate constants can be understood from the amount of delocalization of the respective hydrogen atoms visualized in Fig.~\ref{instpath}. {Note that the pre-reactive complex is very similar to the image of the path where the H atom is the furthest away from the \ce{H2S} molecule. Other pre-reactive complexes may also be found, \emph{e.g.}, where the incoming H-atom interacts with the S-atom of the \ce{H2S} molecule directly. The difference in the interaction energies is small, $\sim0.2$~kJ/mol, and therefore the resulting rate constants are very similar as well. } Substituting the incoming atom as well as the one to be abstracted leads to an order of magnitude difference in the rate constant at low temperature, whereas the exchange of the spectator atom has only a vanishing influence. 

\begin{table}
  \caption{Unimolecular vibrationally adiabatic barriers and crossover temperatures with respect to the encounter complex}
  \label{unimolbar}
  \begin{tabular}{l@{}lcc}
    \hline
    Reaction	&	& $E_\text{uni, act.}$ (kJ/mol) & $T_\text{c}$ (K)\\
    \hline
    \ce{H + H2S} & \ce{ -> H2 + HS} & 12.6 & 286 \\
    \ce{H + HDS} & \ce{ -> H2 + DS} & 12.5 & 286 \\
    \ce{H + HDS} & \ce{ -> HD + HS} & 15.0 & 246 \\
    \ce{H + D2S} & \ce{ -> HD + DS} & 15.0 & 246 \\
    \ce{D + H2S} & \ce{ -> HD + HS} & 11.3 & 234 \\
    \ce{D + HDS} & \ce{ -> HD + DS} & 11.2 & 234 \\
    \ce{D + HDS} & \ce{ -> D2 + HS} & 13.7 & 202 \\
    \ce{D + D2S} & \ce{ -> D2 + DS} & 13.7 & 202 \\
    \hline
  \end{tabular}
\end{table}

All values for the rate constants are listed in the Supporting Information and can thus be used in future astrochemical models. For extrapolation to 10~K we propose to use the values at 55~K for the H-addition reactions as the rate constants calculated at 70~K and 55~K are nearly identical. As the D-addition reactions have not yet fully reached an asymptotic value at 55~K, these rate constants can be best seen as upper limits. Unfortunately, it is not possible to calculate rate constants at lower temperatures, because the number of images needs to be increased significantly which increases the computational cost considerably.

\subsection{Influence of surface molecules}
For the reverse reaction where the sulfur atom is replaced by an oxygen atom, \ce{H2 + OH -> H + H2O}, \citet{Meisner:2017} showed that a water surface had only a very small influence on the reaction. This leads us to believe that there may not be a significant influence of spectator molecules for this reaction as well. To test this hypothesis various water and hydrogen sulfide clusters are constructed to validate the use of the implicit surface model. Table~\ref{cluster} summarizes the difference in activation energy with respect to the encounter complex between the title reaction in the gas phase and on a cluster. These results are calculated fully on the MPWB1K/def2-TZVP level. The differences range between $-$1.8 and +1.2 kJ/mol with one additional value at 4.1 kJ/mol. In the latter case, the hydrogen atom is oriented such that the distance between the incoming H atom and the closest O-atom is only 2.18 \AA. Such an orientation was, however, not obtained in any of the other clusters where the H--O distance amounts to $\sim$2.7~\AA. Short H--S distances were also found for the \ce{H2S} cluster with 5 molecules. The interaction between H and S is, however, not as strong as for the O atom and as such it does not influence the activation energy to a large extent. In conclusion, this approach indeed shows that the influence of neighboring molecules on the reaction is minimal. Therefore, the rate constants calculated within the implicit surface model are a good representation of the average rate constant that may be expected on a surface. The Cartesian coordinates of all transition state structures for the clusters discussed here are given in the Supporting Information along with an image for each structure.

\begin{table}
  \caption{Energy differences in kJ/mol of $E_\text{uni, act.}$ for the title reaction between taking place on a molecular cluster and in the gas phase}
  \label{cluster}
  \begin{tabular}{lrllr}
    \hline
    Cluster type	& $\Delta E_\text{act.}$ &&  Cluster type	& $\Delta E_\text{act.}$ \\
    \hline
    2 \ce{H2O} A 	& -0.2 & &  2 \ce{H2S} A 	& +0.7 \\
    2 \ce{H2O} B 	& +4.1 & &  2 \ce{H2S} B 	& -1.8 \\
    4 \ce{H2O} A 	& -0.0 & &  5 \ce{H2S} A 	& -1.5 \\    
    4 \ce{H2O} B 	& +1.2 & &  5 \ce{H2S} B 	& +0.5 \\        
    5 \ce{H2O} A 	& -1.6 & &  5 \ce{H2S} C 	& -1.4 \\    
    5 \ce{H2O} B 	& -1.1 & & \\        
 
    \hline
  \end{tabular}\\
\end{table}

\section{Conclusion}

We have studied the kinetics of the reaction \ce{H + H2S -> H2 + HS} with the aim of providing the astrochemical community with accurate rate constants {relevant to both gas-phase and surface reactions} down to unprecedentedly low temperatures. {Therefore, both bi- and unimolecular rate constants are calculated at the  UCCSD(T)-F12/cc-VTZ-F12 level, while the main focus lies on surface reactions that take place following a Langmuir-Hinshelwood mechanism.}  The corresponding kinetic isotope effects are also presented, all within the implicit surface model approximation. The use of this approximation has been validated by investigating the effect of small molecular clusters, up to five molecules, on the energetics of the reaction at the MPWB1K/def2-TZVP level of theory. This has indicated that indeed for the title reaction, the surrounding molecules affect the activation energy only to a small extent. Hence, the rate constants calculated are assumed to be representative for the average value that may be expected for the reaction when it takes place on the surface of an icy interstellar dust grain.

\begin{acknowledgement}

The authors acknowledge support for computer time by the state of Baden-W\"{u}rttemberg through bwHPC and the Germany Research Foundation (DFG) through grant no. INST 40/467-1FUGG. 
This project was financially supported by the European Union's Horizon 2020 research and innovation programme (grant agreement No. 646717, TUNNELCHEM). T.L. wishes to acknowledge the Alexander von Humboldt Foundation for generous support.  We thank Sonia \'Alvarez-Barcia, Jan Meisner, Yasuhiro Oba, Takuto Tomaru, and Naoki Watanabe for useful discussions. 

\end{acknowledgement}

\begin{suppinfo}

\begin{itemize}
  \item Cluster Geometries : images and coordinates of the various transition state geometries calculated on \ce{H2O} and \ce{H2S} molecular clusters with MPWB1K/def2-TZVP.
  \item Rate Constants: collection of tables with 1) bimolecular rate constants in the temperature range between 2000 and 70~K calculated with reduced instanton theory above the crossover temperature and with canonical instanton theory below the crossover temperature and 2) unimolecular rate constants in the temperature range 240 to 55~K calculated with canonical instanton theory.
\end{itemize}

\end{suppinfo}

\bibliography{all.bib}

\end{document}